\documentstyle[aps,multicol,graphics,prl,epsf]{revtex}

\begin{document}
\title{Dynamics of Counterion Condensation}
\author{Ramin Golestanian}
\address{
Institute for Theoretical Physics, University of California,
Santa Barbara, CA 93106-4030 \\
and Institute for Advanced Studies in Basic Sciences, Zanjan 45195-159, Iran}
\date{\today}
\maketitle
\begin{abstract}
Using a generalization of the Poisson-Boltzmann equation, dynamics of counterion
condensation is studied. For a single charged plate in the presence of 
counterions, it is shown that the approach to equilibrium is diffusive.
In the far from equilibrium case of a moving charged plate, 
a dynamical counterion condensation transition occurs at a
critical velocity. The complex dynamic behavior of the counterion cloud 
is shown to lead to a novel nonlinear force--velocity relation for the 
moving plate.

\end{abstract}
\begin{multicols}{2}
Many experimental techniques used to probe properties of macroions
involve setting them to motion by applying some external driving mechanism (such
as migration in an applied electric field or hydrodynamic flow, sedimentation, 
etc.), and monitoring the dynamical response. A macroion in solution 
is surrounded by an accompanying cloud of counterions, and the corresponding 
response of the counterions is in general very complex. This often makes 
interpretation of the observed properties of the dynamically perturbed 
system a nontrivial task.

For example, it is known that if a macroion moves, the asymmetry set 
by the motion causes a distortion in the counterion cloud. The distorted cloud, 
in turn, produces a nonzero electric field (called the {\it asymmetry field} 
\cite{Man2}) that opposes the motion of the macroion. This phenomenon
is known as the ``relaxation effect,'' and has been studied in the literature
within the Debye-H\"uckel (DH) theory\cite{LanLif,BJ}.
The approximate DH description, which is based on a linearization of the 
nonlinear Poisson-Boltzmann (PB) equation, is however known to break down when
the macroions are highly charged. In particular, it does not capture
the phenomenon of {\it counterion condensation}, about which a lot has been 
learned by examining the available exact solutions of the full nonlinear 
equilibrium PB equation \cite{BJ}. 

In this Letter, we develop a dynamical generalization of the nonlinear
PB equation, which we use to study the dynamical aspects of counterion 
condensation. For the specific example of a single charged plate in the 
presence of counterions, we show that the equilibrium, corresponding to 
a Gouy-Chapman profile, is approached diffusively. In the far from equilibrium 
case of a moving charged plate we present an exact stationary solution of the 
dynamical PB equation, and show that the cloud of counterions phase separates 
into a comoving part and a condensate of the excess charge which is expelled 
to infinity. The amount of the comoving charge depends on the velocity of the
plate, and is decreased as the velocity increases. At some critical 
velocity $v_c$, the system undergoes a dynamical phase transition, 
and eventually all the counterions evaporate from the neighborhood
of the moving plate. We show that this leads to a novel nonlinear 
force--velocity relation for the moving charged plate in the presence 
of the counterions.

To study the dynamics of counterions in the presence of macroions, one
needs to formulate a dynamical generalization of the PB equation. 
Consider a set of negatively charged macroions described 
by the given charge density $\rho_{m}({\bf x},t)$, embedded in a cloud 
of positively charged counterions with concentration $c({\bf x},t)$\cite{time}.
The electric potential $\phi$({\bf x},t) is then given by the solution of the Poisson
equation:
\begin{equation}	\label{Poi}
-\nabla^2 \phi={4\pi \over \epsilon} (e c+\rho_{m}),
\end{equation}
in which $\epsilon$ is the dielectric constant of the solvent,
and $e$ is the electron charge.
The dynamics of $c({\bf x},t)$ is governed by a continuity equation
of the form $\partial_{t}c+\nabla \cdot {\bf J}=0$, in which
the current density ${\bf J}$ is composed of a deterministic part 
$c {\bf v}$, and a stochastic part $-D \nabla c$ as given
by Fick's law, where ${\bf v}({\bf x},t)$ is the velocity field, and
$D$ is the diffusion constant of the counterions. 
In a mean-field approximation, the velocity field at each point
is determined by the local value of the electric field ${\bf E}({\bf x},t)$
as ${\bf v}=\mu {\bf E}=-\mu \nabla \phi$, where $\mu$ is the electric
mobility of the counterions. Using the self-consistency relation
in the continuity equation, one obtains the so-called Nernst-Planck
equation
\begin{equation}	\label{Con}
\partial_t c=D \nabla^2 c+\mu \nabla \cdot(c \nabla \phi).
\end{equation}
Note that we have simplified the problem by neglecting any couplings to 
the hydrodynamics of the solvent. 
The above two equations (Eqs.(\ref{Poi}) and (\ref{Con})) describe the
nonlinear dynamics of counterions in the presence of a given
distribution of macroions. At equilibrium, $\partial_t c=0$ and
Eq.(\ref{Con}) yields $c \sim \exp(-\mu \phi/D)$, where equilibration
at a temperature $T$ implies an Einstein relation $D=(\mu/e) k_B T$.
Inserting the Boltzmann form for $c$ back into Eq.(\ref{Poi}) then
yields the PB equation. In analogy to the equilibrium case, we can
define a dynamical Bjerrum length $\ell_B=e \mu/\epsilon D$.

The above nonlinear equations are in general very difficult to solve,
which is why they are usually dealt with in a linearized approximation
\cite{LanLif,BJ}. However, the approximation is known to break down in 
the neighborhood of highly charged surfaces, where interesting phenomena
such as counterion condensation take place. To capture the essence
of the nonlinearity in the dynamical context, we restrict ourselves
to a one dimensional case where the equations prove to be more tractable.
Eliminating the concentration field $c$ from the two equations,
one obtains an equation for the electrostatic potential, which can be 
integrated twice to yield
\end{multicols}
\begin{eqnarray}	\label{KPZ}
\partial_t \phi(x,t)-\partial_t \phi(-\infty,t)=D \partial^2_x \phi
+{\mu \over 2}(\partial_x \phi)^2 
+{4\pi \over \epsilon} \int_{-\infty}^x
dx'\left[\mu \rho_m(x',t) \partial_x \phi(x',t)+D \partial_x \rho_m(x',t)
+J_m(x',t)\right],
\end{eqnarray}
\begin{multicols}{2}\noindent
where the macroion current density $J_m$ is obtained from the
continuity equation for the macroions $\partial_t \rho_m+\partial_x J_m=0$,
and the boundary condition $E(-\infty,t)=0$ has been implemented.

The above equation, which can be called a dynamical PB equation \cite{PB}, 
belongs to the general class of the celebrated Kardar-Parisi-Zhang 
(KPZ) equations \cite{KPZ86}, with the coupling to the source
appearing both in the additive and multiplicative forms. 
It is however important to note that in this context the sources represent 
{\it slower} dynamical degrees of freedom corresponding to the motion 
of the macroions (compared to counterions), which makes it somewhat
different from the standard KPZ problems \cite{KPZ86}.

To study how counterions dynamically rearrange themselves around macroions,
we focus on the specific example of a single negatively charged plate
with charge (number) density $\sigma$ and surface area $A$, which is moving at a constant
velocity $v$, {\em i.e.}, $\rho_m(x,t)=-e \sigma \delta(x-vt)$. 
Using a Cole-Hopf transformation 
$\phi(x,t)-\phi(-\infty,t)=2D/\mu \;\ln W(x,t)$ \cite{KPZ86,Brui}, 
Eq.(\ref{KPZ}) can be written as
\begin{eqnarray}	\label{Schr}
\partial_t W&=&D \partial_x^2 W  \nonumber \\
&-& {2 D \over \lambda}\left[\delta(x-vt)+\left({v-\mu E_0 \over D}\right) 
\Theta(x-vt)\right] W,
\end{eqnarray}
where $E_0(t)=-2 D/\mu \; \partial_x \ln W|_{x=vt}$ is the asymmetry 
field, $\Theta(x)$ is the step function, and $\lambda=1/\pi \sigma \ell_B$ 
is a dynamical Gouy-Chapman length. Note that despite its simple form, this 
diffusion-like equation is still nonlinear due to the self-consistent coupling 
of the asymmetry field (electric field at the charged plate).

Let us first assume that the plate is not moving ($v=0$), and ask how the equilibrium
configuration is dynamically approached. If we assume that the initial configuration of 
the counterions is symmetric with respect to the plate, the symmetry will be conserved 
during time evolution rendering $E_0(t)=0$ at all times. This simplifies Eq.(\ref{Schr}) 
to a linear diffusion equation which can be solved exactly. One obtains
\end{multicols}
\begin{equation}	\label{Dif}
{e\phi(x,t) \over k_B T}=2 \ln\left\{1+{|x| \over \lambda}+\int_{-\infty}^{+\infty}
dx' \left[\exp\left({e \phi(x',0) \over 2 k_B T}\right)-1-{|x'| \over \lambda}\right]\times
{1 \over \sqrt{4 \pi D t}} \exp\left[-{(x-x')^2 \over 4 D t}\right]\right\}.
\end{equation}
\begin{multicols}{2}\noindent
In simple terms this means that the equilibrium Gouy-Chapman profile diffusively
develops, as the initial configuration diffusively disappears. Naively, one might
not expect such an amazingly simple behavior from a fully interacting Coulomb system.

For a stationary plate at equilibrium, all the counterions are known to be ``condensed''
in a Gouy-Chapman (density) profile which decays algebraically at large separations
(see below). 
It is interesting to study this problem in the far from equilibrium case of 
a moving plate. In particular, we may ask if there is a stationary solution of 
Eq.(\ref{Schr}), of the form $W(x,t)=W(x-vt)$, in which {\it all} the counterions 
are {\it comoving} with the plate at the velocity $v$. The answer to this question
turns out to be negative. Apparently, the Coulomb attraction which could
overcome the combination of entropy and repulsion among the counterions in the 
equilibrium case, is not capable of accommodating the viscous drag of {\it all} the 
counterions in this balance, for any nonzero $v$. However, we can find 
stationary solutions if we do not require that all the counterions are comoving.

Such solutions can be shown to exist only for negative values of $v$, with a 
boundary condition that requires a specific value for the electric field at $+\infty$.
This implies that the neutralizing counterion cloud actually phase separates into a 
comoving domain and a domain of {\it excess} charges accumulated at $+\infty$, 
as required by this one dimensional model \cite{v<0}. 
This class of solutions is parametrized 
by the number of comoving counterions. The stationary solution of Eq.(\ref{Schr})
with maximum number of comoving counterions can be obtained as \cite{Rava}
\end{multicols}
\begin{equation}
W(x,t)=\left\{\begin{array}{ll}
1-\left({1+{\lambda v/4 D}\over 1-{\lambda v/4 D}}\right)^2 
\exp\left[-{v (x-v t) \over D}\right], & {\rm for} \;\;x < vt,  \\ \\ 
{\left(-\lambda v/D\right) \over (1-{\lambda v/4 D})^2}
\left\{1+\left[1-\left({\lambda v \over 4 D}\right)^2\right] 
{(x-v t) \over \lambda}\right\}
 \exp\left[-{v (x-v t) \over 2 D}\right], & {\rm for} \;\; x > v t,  
\end{array} \right. \label{W1}
\end{equation}
and correspondingly for the counterion density profile as
\begin{equation}
c(x,t)=\left\{\begin{array}{ll}
{(v/D)^2 \over 2 \pi \ell_B}
{\left({1+{\lambda v/4 D}\over 1-{\lambda v/4 D}}\right)^2  
\exp\left[-{v (x-v t) \over D}\right]\over 
\left\{1-\left({1+{\lambda v/4 D}\over 1-{\lambda v/4 D}}\right)^2 
\exp\left[-{v (x-v t) \over D}\right]\right\}^2}, & {\rm for}\;\; x < vt,  \\ \\ 
{1\over 2 \pi \ell_B \lambda^2}{\left[1-\left({\lambda v/4 D}\right)^2\right]^2 
\over \left\{1+\left[1-\left({\lambda v/4 D}\right)^2\right] 
\left({x- v t \over \lambda}\right)\right\}^2}, & {\rm for}\;\; x > v t.  
\end{array} \right. \label{c1}
\end{equation}
\begin{multicols}{2}\noindent

The above solution has very interesting features. In the limit $v \to 0$, 
Eq.(\ref{c1}) reduces to the equilibrium Gouy-Chapman profile \cite{GC}
\begin{equation}	\label{c2}
c_{\rm eq}(x)={1 \over 2\pi \ell_B \lambda^2} \times {1 \over 
\left(1+{|x| \over \lambda}\right)^2}.
\end{equation}
In this limit, all the counterions are condensed, typically at a distance
given by $\lambda$, although the profile has an algebraic decay.  

For a nonzero $v$, while the density profile
is still algebraically decaying ``behind'' the moving plate (for $x > v t$),
it decays exponentially ``ahead'' of the moving plate (for $x < v t$),
indicating a {\it jamming} of the profile as a result of the motion. In the
comoving profile of counterions, there is a total charge of 
$Q_{>}={1 \over 2} Q_0 [1-(\lambda v/4 D)^2]$ behind the plate, as opposed to 
$Q_{<}={1 \over 2} Q_0 [1+(\lambda v/4 D)^2+\lambda v/2 D]$ ahead
of the plate, where $Q_0=e \sigma A$ is the overall charge of the counterions.
Note that the comoving counterions are distributed asymmetrically with 
$Q_{>} > Q_{<}$. The amount of the excess charge 
can be easily obtained as 
\begin{equation}
Q_{\rm ex}(v)=Q_0-Q_{>}-Q_{<}={e A \over 4 \pi \ell_B} \times {|v| \over D},
\label{Qex}
\end{equation}
which is independent of $\sigma$. By examining the asymptotic limit of the
electric field at infinity, one can easily see that this excess charge 
is indeed expelled to $+\infty$.

As the velocity of the plate increases, the number of comoving counterions 
decreases, until at $\lambda v/4 D=-1$ eventually all the counterions 
evaporate to infinity. This corresponds to a {\it dynamical} counterion 
condensation phase transition, happening at the critical velocity 
$v_c=4 D/\lambda$, and is similar to the equilibrium transition in the
case of a charged cylinder \cite{Man,Oos}. 
It is easy to check from Eq.(\ref{c1}) that the 
density profile of the comoving counterions vanishes on both sides at 
$\lambda v/4 D=-1$.

The solution of Eq.(\ref{W1}) yields a value $E_0=(1+\lambda v/8 D)v/\mu$
for the asymmetry field at the position of the plate, which leads to a very 
interesting mechanical response for the charged plate \cite{E0}. 
In the stationary situation, the total force $F_{\rm tot}$ exerted on the 
charged plate should vanish. This means that an externally applied 
(mechanical) force $F_{\rm ext}$ (that is necessary to maintain the constant
velocity motion) should balance an electrical contribution 
due to nonzero value of the electric field $E_0$ felt by the charged plate, 
namely, $F_{\rm ext}-Q_0 E_0=0$.
Inserting the velocity dependent value for $E_0$, one obtains
\begin{equation}
|F_{\rm ext}|=\left\{\begin{array}{ll}
\left({Q_0/\mu}\right) |v|-{\epsilon \over 8 \pi}
\left(v/\mu \right)^2 A, & {\rm for}\;\; |v| < v_c,  \\ \\
2 \pi Q_0^2/\epsilon A, & {\rm for}\;\; |v| \geq v_c,  
\end{array} \right. \label{Fv1}
\end{equation}
as the force--velocity relation for a moving charged plate. 
A plot of Eq.(\ref{Fv1}) is sketched in Fig.~1, in which
$F_c=2 \pi Q_0^2/\epsilon A$.

\begin{figure}
\centerline{\epsfxsize 6cm \rotatebox{-90}{\epsffile{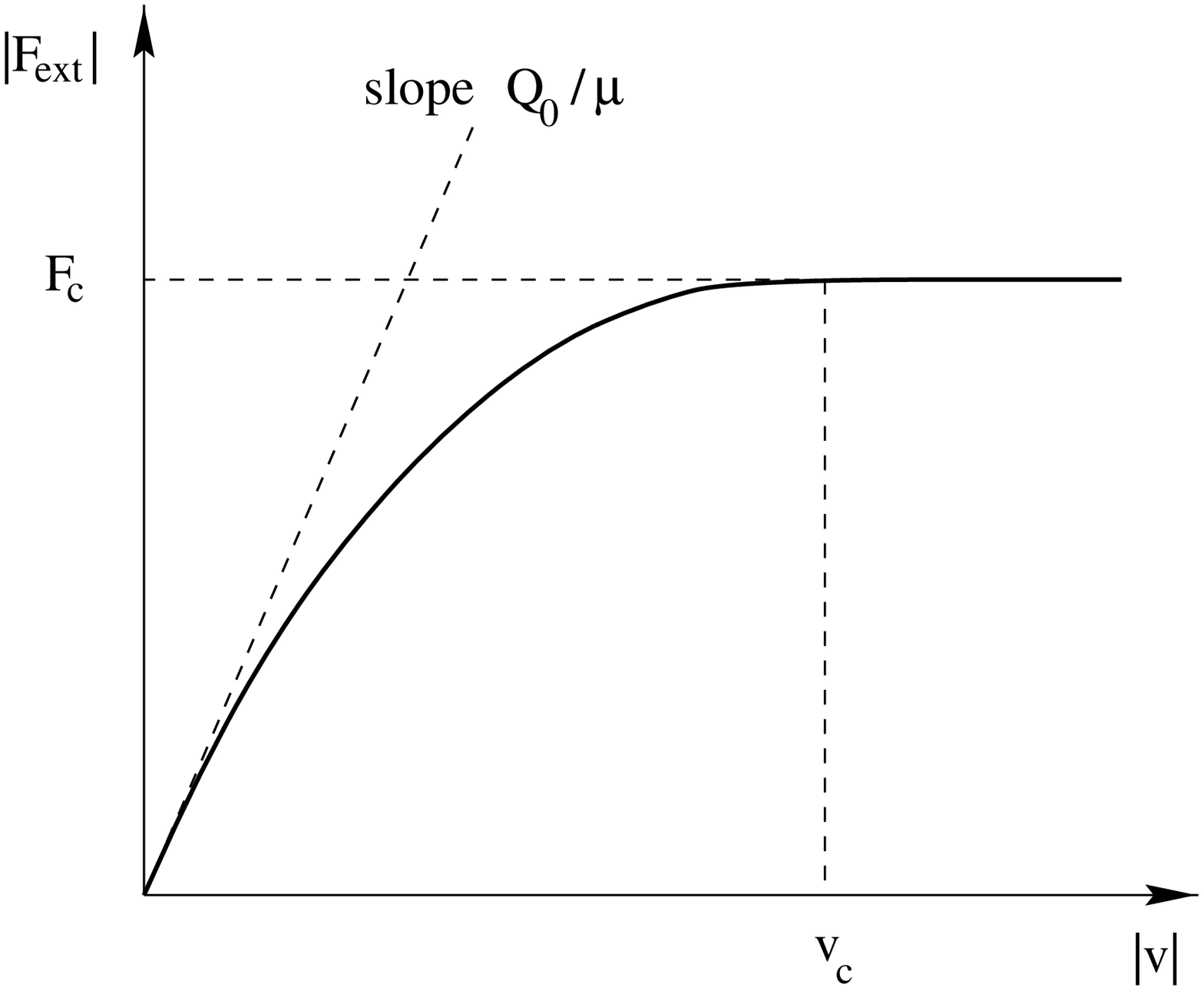}}}\vskip .3truecm
FIG.~1. A plot of nonlinear force--velocity curve for the moving charged plate.
The slope of the curve at any velocity is a measure of the number of condensed
(comoving) counterions.
\label{fig1}\end{figure}

In the limit of very low velocities ($|v| \ll v_c$), one can neglect the 
quadratic velocity term in Eq.(\ref{Fv1}). This leads to a viscous
drag with an effective friction coefficient $\zeta_{\rm eff}=Q_0/\mu$,
where the ``electrostatic friction'' is due to dissipation corresponding
to the comoving could of counterions. Note that all the counterions
are condensed in this limit, and participate in the dissipation.
In the opposite limit of $|v| \gg v_c$, no counterions are condensed
and the friction coefficient is consequently zero. However,
since all the counterions participate in forming a condensate at infinity,
there is a constant (independent of velocity) electrostatic attractive
force acting on the plate, which is simply the force between two parallel
and equally charged plates. In fact, one can rewrite Eq.(\ref{Fv1}) in such a
way that these features are manifest at any velocity:
$|F_{\rm ext}|=(Q_c(v)/\mu) |v|+2\pi Q_{\rm ex}(v)^2/\epsilon A$, where
$Q_c(v)=Q_{>}+Q_{<}=Q_0-Q_{\rm ex}(v)$ is the amount of condensed (comoving) 
charge. Interestingly, the nonlinear response function 
\begin{equation}	\label{Zeta}
\zeta (v) \equiv {\partial F_{\rm ext}(v) \over \partial v}=Q_c(v)/\mu,
\end{equation}
gives a proper account of the number of condensed (comoving) counterions at any 
velocity. 

The dynamical counterion condensation phase transition discussed above
can be characterized by an order parameter, which is given by the 
number of condensed counterions. The tunning parameter for this
dynamical transition is the external force, which is the analogue of 
temperature in the equilibrium case. In analogy to equilibrium critical
phenomena, one can look at the critical exponents corresponding to the 
transition. For example, the order parameter vanishes at $F_c$ as 
$Q_c \sim (F_c-|F_{\rm ext}|)^{\beta}$, with a mean-field exponent $\beta=1/2$.

It is well known that at equilibrium counterion condensation is determined
by dimensional considerations \cite{Zimm}. In $d$ dimensions, a $D$ dimensional
macroion of size $L_{\parallel}$ attracts a counterion at a perpendicular
distance $L_{\perp}$, with a Coulomb energy that goes like 
$E_c \sim 1/L_{\perp}^{d-D-2}$. On the other hand, entropy
of a particle confined in such a box can be obtained as 
$S \sim \ln(L_{\parallel}^D L_{\perp}^{d-D})$. Comparing energy and 
entropy one obtains that for $d<D+2$ counterions tend to condense
to minimize energy (lower values for $L_{\perp}$ are favored), 
while for $d>D+2$ they prefer to be free and
gain entropy (higher values for $L_{\perp}$ are favored). 
The case $d=D+2$ is marginal, where a counterion condensation
transition takes place \cite{Man,Oos}. The analysis presented in this
Letter clearly shows that this simple equilibrium picture is modified
when the system is far from equilibrium. In particular, we found that
complete condensation in the case of a charged plate will no longer
hold for a moving plate. Another important difference is that, unlike 
the equilibrium case, there is a marked contrast between the 
condensed and the free counterions. 

One might question the validity of the present formulation at high
velocities such that $\lambda v/D \sim 1$, because of the assumption
that the dynamics of the macroions is much slower compared to the counterions.
The dimensionless parameter $\lambda v/D$ (often called the P\'eclet number) 
can be written as $\tau_c/\tau_m$, 
in which $\tau_c$ is the time it takes a counterion to {\it diffuse} a
distance $\lambda$, whereas $\tau_m$ is the time in which the macroion 
{\it drifts} a same distance. It is thus possible that this ratio becomes 
comparable to 1, while the ratio between the corresponding diffusion times 
is still much less than unity.

In the derivation of Eq.(\ref{Con}), we have made use of a mean-field
approximation similar to what is used to obtain the PB equation.
It is well known that PB equation is incapable of describing intriguing
phenomena such as attraction between like charged objects in the presence
of condensed counterions, because of this approximation \cite{attract}.
One should then attempt to go beyond mean-field theory and study 
the effect of fluctuations on the dynamical PB equation along the 
lines of what has been done in the equilibrium case
\cite{Pod,KGRMP,NO}. In particular, it is shown in Ref.
\cite{NO} that fluctuations lead to a stronger condensation, because 
mean-field theory overestimates the repulsion between counterions.
It will thus be interesting to see how this effect competes with the
viscous drag in the case of a moving plate.
Another important and very interesting extension of this work would be
to incorporate the coupling to hydrodynamics of the solvent, which 
we anticipate to have dramatic effects.

I am grateful to R. Bruinsma, M. Kardar, P. Pincus, M. Sahimi, and
R. da Silveira for invaluable discussions and comments. This research was 
supported in part by the National Science Foundation under Grants 
No. PHY94-07194 and DMR-93-03667.

\end{multicols}

\begin{references}
\bibitem{time}
We have assumed that the dynamics of the macroions is considerably slower
than that of the counterions.
\bibitem{PB}
Note that Eq.(\ref{KPZ}) gives an alternative form for the equilibrium PB 
equation by setting $\partial_t \phi=0$, with a quadratic (rather than 
exponential) nonlinearity.
\bibitem{v<0}
The fact that no solutions exist for positive values of $v$ is because
we have imposed a zero electric field boundary condition at $-\infty$ in the
derivation of Eq.(\ref{KPZ}), which does not allow the excess charges
to form a condensate at $-\infty$.
\bibitem{Rava}
One can show that the solution with maximum number of comoving
counterions is the stable solution. I am indebted to R. da Silveira
for pointing this out to me.
\end{references}
\end{document}